\documentclass[12pt]{article}

\usepackage[latin1]{inputenc}

\usepackage{graphicx,xcolor,%
natbib,amssymb,amsmath}

\usepackage{euscript}

\newtheorem{theoreme}{Theorem}
\newtheorem{lemma}[theoreme]{Lemma}
\newtheorem{proposition}[theoreme]{Proposition}

\newenvironment{proof}{\small{\bf Proof.}}{\hfill$\Box$\normalsize
\bigskip}

\newcommand{\actes}{{\mathbb A}} %"ensemble des reels"
\newcommand{\acte}{{\mathtt a}}
\newcommand{\RR}{{\mathbb R}} %"ensemble des reels"
\newcommand{\EE}{{\mathbb E}} %"symbole d'esperance mathematique"
\newcommand{\PP}{{\mathbb P}} %"symbole de probabilite"
\newcommand{\Risk}{{\mathcal R}} %"fonction caractéristique"

\def\euro{\mbox{\raisebox{.25ex}{{\it =}}\hspace{-.5em}{\sf C}}}

\def\text#1{\quad\mbox{#1}\quad} %wide text in mathsok
\def\defegal{:=} 

\def\euro{\mbox{\raisebox{.25ex}{{\it =}}\hspace{-.5em}{\sf C}}} %Commande Euro

\def\cvar{\mathsf{CVaR}}

\def\VaR{\mathsf{VaR}}
\def\Wmd{\mathsf{WMd}}
\def\util{U}
\def\wealth{w}

\newcommand{\vari}{\mathop{\mathsf{var}}}
\def\return{J}

\def\opt{^\sharp}
\newenvironment{preuve}{\small{\bf
    Proof:}}{\hfill$\Box$\normalsize\\ \bigskip}

\def\mtext#1{\,\mbox{#1}\,} %text in maths

\title{Conditional Value-at-Risk Constraint\\
 and Loss Aversion Utility Functions}

\author{Laetitia Andrieu\footnote{
Électricité de France R\&D,
 1 avenue du Général de Gaulle,
F-92141 Clamart Cedex, France,  Email: laetitia.andrieu@edf.fr
},
Michel De Lara\footnote{
Université Paris--Est, \textsc{CERMICS}, 
 6--8 avenue Blaise Pascal, 77455 Marne la Vallée Cedex 2, France. 
Corresponding author:  delara@cermics.enpc.fr, fax +33164153586
},
Babacar Seck\footnote{
Université Paris--Est, \textsc{CERMICS}, 
 6--8 avenue Blaise Pascal, 77455 Marne la Vallée Cedex 2, France. 
Email: seck@cermics.enpc.fr
}
}

\begin{document}
\maketitle

%\setcounter{tocdepth}{1} %LC-35
%\tableofcontents

\begin{abstract}
We provide an economic interpretation of the practice
consisting in incorporating risk measures as constraints in a
classic expected return maximization problem.
For what we call the infimum of expectations class of risk measures,
we show that if the decision maker (DM) maximizes the
expectation of a random return under constraint that the risk measure is
bounded above, he then behaves as a
``generalized expected utility maximizer'' in the following sense. 
The DM exhibits ambiguity with respect to a family of utility
functions defined on a larger set of decisions than the original one;
he adopts pessimism and performs first a minimization of expected utility
over this family, then performs a maximization over a new decisions set.
This economic behaviour is called ``Maxmin under risk'' and studied by
Maccheroni (2002).
This economic interpretation allows us to exhibit a loss aversion factor
when the risk measure is the Conditional Value-at-Risk.
\bigskip


\end{abstract}

%\textbf{JEL Classification:}  D81
% JEL : D81 

\section{Motivation}

Taking risk into account in decision problems in a mathematical formal
way is more and more widespread. For instance, liberalization of energy 
markets displays new issues for electrical
companies which now have to master
both traditional problems (such as optimization of electrical generation)
and emerging problems (such as integration of spot markets and risk
management, see e.g.~\cite{Romisch:risk}). The historical
issue which consisted in managing the electrical generation at
lowest cost evolved: liberalization of energy markets and
introduction of spot markets lead to consider a problem of revenue
maximization  under Earning-at-Risk constraint, because financial risks
are now added to the traditional risks.

Let us now be slightly more formal.
Consider a decision maker (DM)
whose return $\return(\acte,\xi)$ depends on a decision variable
$\acte$ (for instance, proportions of assets) and a random variable
$\xi$ (random return of assets, for example).
The question of how to take risk into account in addition has been
studied since long.
Let us briefly describe two classical approaches to deal with risk in a
decision problem. On the one hand,
the DM may maximize the expectation of $\return(\acte,\xi)$ under explicit
risk constraints, such as variance (as in~\cite{Markowitz:1952})
or Conditional Value-at-Risk; we shall coin this practice
as belonging to the engineers or practitioners world.
On the other hand, the DM may maximize
the expectation of $\util\big(\return(\acte,\xi)\big)$ where $\util$ is a
utility function wich captures more or less risk aversion (in the so
called expected utility theory, or more general functionals else);
this is the world of economists.
In this paper, we shall focus on the links between these two approaches.

The paper is organized as follows. In Section~\ref{sec:Risk-as-infimum},
we specify a wide class of risk measures which will prove
useful to provide an economic interpretation of profit
maximization under risk constraint. Section~\ref{sec:Main-result}
gives our main result and points out a specific 
nonexpected utility theory which is compatible with the original problem.
Our approach uses duality theory and Lagrange multipliers.
However it does not focus on the optimal multiplier, and this is how we
obtain a \emph{family} of utility functions and an economic interpretation
(though belonging to nonexpected utility theories),
and not a single  utility function.
This differs from the result in~\cite{DDAR06:risque} 
where the authors
prove, in a way, that utility functions play the role of
Lagrange multipliers for second order stochastic dominance constraints.
With this, \cite{DDAR06:risque} prove the equivalence between
portfolio maximization under second order stochastic dominance constraints
and expected utility maximization, for \emph{one single} utility function.
However, such utility function is not given \emph{a priori} and
may not be interpreted economically before the decision problem.
A concluding economic discussion is given in Section~\ref{sec:Conclusion}.
Proofs are gathered in Appendix~\ref{sec:Appendix}.

\section{The infimum of expectations class of risk measures}
\label{sec:Risk-as-infimum}

Let be given a probability space $(\Omega,\mathcal{F},\PP)$ with $\PP$
a probability measure on a $\sigma$-field
$\mathcal{F}$ of events on $\Omega$.
We are thus in a \emph{risk} decision context.
The expectation of a random variable on $(\Omega,\mathcal{F},\PP)$ will
be denoted by $\EE$.
We introduce a class of risk measures which covers many of the
usual risk measures and which will be prove adequate for optimization 
problems. We will also briefly examine the connections with coherent
risk measures.

\subsection{Definition and examples}
\label{ss:Risk-as-infimum}

Let a function $\rho : \RR\times\RR \to \RR $ be given.
Let $L_{\rho}(\Omega,\mathcal{F},\PP)$ be a set of
random variables $X$ defined on $(\Omega,\mathcal{F},\PP)$ such that
$\rho\big(X,\eta\big)$ is integrable for all $\eta$
and such that the following risk measure
\begin{equation}
\Risk_{\rho}(X) \defegal
\inf_{\eta\in\RR}\EE\big[\rho\big(X,\eta\big)\big]\; ,
\label{eq:Risk-formulation}
\end{equation}
is finite ($\Risk_{\rho}(X) > -\infty$).

In the sequel, we shall require the following properties for the
function $\rho : \RR\times\RR \to \RR $ and for
$L_{\rho}(\Omega,\mathcal{F},\PP)$.
\begin{enumerate}
\item[H1.] $\eta \mapsto  \rho(x,\eta)$ is a convex function,
\item[H2.] for all $X \in L_{\rho}(\Omega,\mathcal{F},\PP)$ ,
$\eta \mapsto \EE\big[\rho\big(X,\eta\big)\big]$
is continuous\footnote{%
We could also formulate assumptions directly on the function $\rho$, 
ensuring Lebesgue dominated convergence validity, 
but this is not our main concern here.} 
and has limit $+\infty$ when $\eta \to +\infty$.
\end{enumerate}

The random variable $X$ represents a loss. Hence,
risk constraints will be of the form $\Risk_{\rho}(X) \leq \gamma$.
For the so called safety measures, the safety constraint is rather
${\cal S}_{\rho}(X) \geq \gamma$. We pass from one to the other by
$\displaystyle{\cal S}_{\rho}(X) = - \Risk_{\rho}(-X) =
\sup_{\eta\in\RR}\EE\big[-\rho\big(-X,\eta\big)\big]$.

Several well-known risk measures belong to the
infimum of expectations class of risk measures.

\subsubsection*{Variance}

In~\cite{Markowitz:1952}, Markowitz  uses variance as a risk 
measure.
A well known formula for the variance is
\begin{equation*}
\vari\big[X\big]=\inf_{\eta\in\RR}
\EE\big[\big(X-\eta\big)^2\big]\,.
\end{equation*}
The function
\begin{equation*}
\rho_{\vari}(x,\eta):=(x-\eta)^{2}
\end{equation*}
is convex with respect to $\eta$ (H1.).
Taking $L_{\rho}(\Omega,\mathcal{F},\PP)=L^2(\Omega,\mathcal{F},\PP)$,
assumption H2. is satisfied.

\subsubsection*{Conditional Value-at-Risk}

\cite{RockUry:risk} give the following formula for the 
Conditional Value-at-Risk risk measure $\cvar$ at confidence
level\footnote{% 
In practice, $p$ is rather close to 1 ($p=0.95$, $p=0.99$).
The Value-at-Risk 
$\VaR_{p}(X)$ is such that $\PP(X \leq \VaR_{p}(X)) = p$. 
Then $\cvar_{p}(X) $ is interpreted as the expected value of $X$ 
knowing that $X$ exceeds $\VaR_{p}(X)$.
} 
$0<p<1$:
\begin{equation*}
\cvar_{p}(X) = \inf_{\eta \in \RR} \left(
\eta+\frac{1}{1-p}\,\EE \big{\lbrack} \max \lbrace0, X -\eta \rbrace
\big{\rbrack} \right) \, .
\end{equation*}
The function
\begin{equation*}
\rho_{\cvar}(x,\eta) \defegal \eta+\frac{1}{1-p}
\max \lbrace0, x -\eta \rbrace
\end{equation*}
is convex with respect to $\eta$.
Taking $L_{\rho}(\Omega,\mathcal{F},\PP)=L^1(\Omega,\mathcal{F},\PP)$,
assumption H2. is satisfied.

\subsubsection*{Weighted mean deviation from a quantile}

Let us introduce $\psi_{X}^{-1}$ the left-continuous inverse of the
cumulative distribution function $\psi_{X}$ of the random variable $X$
and $\psi_{X}^{-2}(p) = \int_{0}^{p} \psi_{X}^{-1}(\alpha)\mathrm{d}\alpha$.
The weighted mean deviation from a quantile $\Wmd$ is
\begin{equation*}
\Wmd_{p}(X)=\EE[X]\,p-\psi_{X}^{-2}(p)\,.
\end{equation*}
In \cite{Rus3:1999}, one finds the expression
\begin{equation*}
\Wmd_{p}(X)= \inf_{\eta \in \mathbb{R}} \EE \Big{\lbrack}
\max\big\{p(X-\eta),(1-p)(\eta-X)\big\} \Big{\rbrack} \, .
\end{equation*}
The function
\begin{equation*}
\rho_{\Wmd}(x,\eta):=\max\big\{p(x-\eta),(1-p)(\eta-x)\big\}\, ,
\end{equation*}
is continuous, convex with respect to $\eta$.

\subsubsection*{Optimized Certainty Equivalent}

The Optimized Certainty Equivalent was introduced in
\cite{Teboulle:util} (see also \cite{Teboulle:util2}).
This concept is based on the economic notion of
certainty equivalent. Let $\util:\RR\to[-\infty,+\infty[$ be
a proper closed  concave and nondecreasing
utility function\footnote{%
With additional assumptions:
$\util$ has effective domain $\mathrm{dom} \util=\{x\in\RR\mid
\util(x)>-\infty \}\neq\emptyset $, supposed to satisfy
$\util(0)=0$ and $1 \in\partial\util(0)$,
where $\partial\util$ denotes the subdifferential map of $\util$.}. The
Optimized Certainty Equivalent ${\cal S}_{\util} $ of the random return
$X$ defined on $(\Omega,\mathcal{F},\PP)$ is
\begin{equation*}
{\cal S}_{\util}(X) = \sup_{\nu \in \RR} \left(
\nu + \EE\big[\util(X-\nu)\big] \right).
\end{equation*}
The associated risk measure is
\begin{equation*}
\Risk_{\rho}(X) \defegal -{\cal S}_{\util}(-X)  =
\inf_{\eta \in \RR} \left(
\eta - \EE\big[\util(\eta-X)\big] \right) 
\end{equation*}
where 
\begin{equation*}
\rho_{\util}(x,\eta)\defegal \eta - \util(\eta-x)
\end{equation*}
satisfies assumptions H1. and H2. whenever $\util$ satisfies
appropriate continuity and growth assumptions.

\subsubsection*{A Summary Table}

We sum up the above cases. We denote $a_+ :=\max\{a,0\}$.
Notice that all the functions $x \mapsto \rho(x,\eta)$ in
Table~\ref{tab:Risk_measures_given_by_an_infimum_of_expectations} are
convex. 

\begin{table}[htbp]
\begin{center}
\begin{tabular}{c||c}
\hline
Risk measure $\Risk_{\rho}$ & $\rho(x,\eta)$ \\[2mm]
\hline\hline
                  &                           \\
\emph{Variance}          &  $(x-\eta)^2$     \\[2mm]
\emph{Conditional Value-at-Risk}  &  $\eta+\frac{1}{1-p}(x-\eta)_+$ \\[3mm]
\emph{Weighted Mean Deviation} &
$ \max\big\{p(x-\eta),(1-p)(\eta-x)\big\}$ \\[3mm]
\emph{Optimized Certainty Equivalent} &
$\eta - {\util}(\eta-x) $ \\
\hline
\end{tabular}
\end{center}
\caption{Risk measures given by an infimum of expectations}
\label{tab:Risk_measures_given_by_an_infimum_of_expectations}
\end{table}

\subsection{Infimum of expectations and coherent risk measures}

Coherent risk measures were introduced in \cite{Delbaen:1999}.
When the risk measure $\Risk_{\rho}$ is given by \eqref{eq:Risk-formulation},
we shall provide sufficient conditions on $\rho$ ensuring monotonicity,
translation invariance, positive homogeneity, convexity,
subadditivity.

\begin{proposition}
Under the assumption that the set $L_{\rho}(\Omega,\mathcal{F},\PP)$ in
\S\ref{ss:Risk-as-infimum} is a vector space containing the constant
random variables, we have the following properties.

\begin{enumerate}
\item If $x \mapsto  \rho(x,\eta)$ is increasing, then $\Risk_{\rho}$
is monotonous.

\item If $\rho(x+m,\eta)=\rho(x,\eta'_{m})-m$ where $\eta \mapsto \eta'_{m}$ is
  one-to-one, then $\Risk_{\rho}$ satisfies translation invariance.

\item If $\rho(\theta x,\eta)=\theta \rho(x,\eta'_{\theta})$
where $\eta \mapsto \eta'_{\theta}$ is
  one-to-one for $\theta >0$,
then $\Risk_{\rho}$ satisfies positive homogeneity.

\item If $(x,\eta) \mapsto  \rho(x,\eta)$ is jointly convex
then $\Risk_{\rho}$ 
      is convex.
\label{it:convex}

\item If $(x,\eta) \mapsto  \rho(x,\eta)$ is jointly subadditive,
then $\Risk_{\rho}$ is subadditive.
\label{it:subadditive}
\end{enumerate}

\end{proposition}

\begin{preuve}
We shall only prove item~\ref{it:convex} since item~\ref{it:subadditive}
is proved in the same way, and that all other assertions are straightforward.

Assume that $(x,\eta) \mapsto  \rho(x,\eta)$ is jointly convex.
Let $X_1$ and $X_2$ be two random variables,
$(\eta_1,\eta_2)\in \RR^2$ and $\theta \in[0,1]$.
We have
\[
\rho(\theta X_1 + (1-\theta)X_2,\theta \eta_1 + (1-\theta)\eta_2)
\leq \theta\rho(X_{1},\eta_1) + (1-\theta)\rho(X_{2},\eta_2).
\]
By using expectation operator and positivity of $\theta$, one obtains
\[
\inf_{\eta}\EE\big[\rho(\theta X_1 + (1-\theta)X_2,\eta)\big]\leq
\inf_{(\eta_1,\eta_2)\in\RR\times\RR}
\Big\{\theta\EE\big[\rho(X_{1},\eta_1)\big] +
(1-\theta)\EE\big[\rho(X_{2},\eta_2)\big]\Big\}.
\]
It yields
\[
\Risk_{\rho}(\theta X_1 + (1-\theta)X_2)\leq
\theta\Risk_{\rho}(X_1)+(1-\theta)\Risk_{\rho}(X_2) \; .
\]
\end{preuve}

For the Conditional Value-at-Risk, $\rho_{\cvar}(x+m,\eta) =
\rho_{\cvar}(x,\eta'_{m})-m$ where $\eta'_{m}=\eta +m$.
For the Weighted Mean Deviation, $\rho_{\Wmd}(\theta x,\eta)=
\theta \rho(x,\eta')$ where $\eta'_{\theta} = \eta / \theta$.

\section{Profit maximization under risk constraints: a reformulation
  with utility functions}
\label{sec:Main-result}
 
We now state our main result, which is an equivalence between a
profit maximization under risk constraints problem
and a maxmin problem involving an infinite number of utility
functions.
We relate this (non) expected utility economic interpretation
to the ``maxmin'' representation proposed in~\cite{Maccheroni:2002}.
For the specific case of $\cvar$  risk constraint, we exhibit a link
with ``loss aversion utility functions'' \emph{à la} Kahneman and Tversky 
(see~\cite{KT92}).

\subsection{A maxmin reformulation}
To formulate a maximization problem under risk constraint, let us
introduce
\begin{itemize}
\item a set $\actes \subset \RR^n$ of actions or decisions, 
\item  a random variable $\xi$ defined on a probability space
$(\Omega,\mathcal{F},\PP)$, with values in $\RR^p$,
\item a mapping $\return : \RR^n \times \RR^p \to \RR $,
such that for any action $\acte \in \actes$, the random variable
$ \return(\acte,\xi)$ represents the prospect
(profit, benefit, etc.) of the decision maker,
\item a risk measure $\Risk_{\rho}$ defined in~\eqref{eq:Risk-formulation},
together with the level constraint $\gamma\in\RR$.
\end{itemize}

Our main result is the following (the proof is given in
Appendix~\ref{sec:Appendix}). 

\begin{theoreme}
Assume that $\rho$ and $L_{\rho}(\Omega,\mathcal{F},\PP)$
satisfy assumptions H1. and H2. in~\S\ref{ss:Risk-as-infimum},
that $ \return(\acte,\xi) \in L_{\rho}(\Omega,\mathcal{F},\PP)$
for all $\acte \in \actes$,
and that the infimum in~\eqref{eq:Risk-formulation} is achieved for any
loss $X=-\return(\acte,\xi)$ when $\acte$ varies in $\actes$.

The maximization under risk constraint\footnote{%
The risk constraint is not on the prospect $\return(\acte,\xi)$, but on 
the loss $-\return(\acte,\xi)$.
\label{ft:risk_loss}}
problem 

\begin{equation}
\left\{ \begin{array}{rcl}
\displaystyle\sup_{\acte\in\actes} \EE \big[\return(\acte,\xi)\big]\\[4mm]
\Risk_{\rho}\big(-\return(\acte,\xi)\big) =
\displaystyle\inf_{\eta\in\RR}
\EE\big[\rho\big(-\return(\acte,\xi),\eta\big)\big] \leq \gamma,
\end{array}\right.
\label{eq:Pb-ref}
\end{equation}
is equivalent to the following maxmin problem 
\begin{equation}
\label{eq:Eco}
\sup_{(\acte,\eta)\in\actes\times\RR}\inf_{U\in\mathcal{U}}
\EE \left[ \util \big(\return(\acte,\xi),\eta\big) \right] \; .
\end{equation}
The set $\mathcal{U}$ of functions $\RR^2 \to \RR$
over which the infimum is taken is 
\begin{equation*}
%\label{eq:Utility-ref}
\mathcal{U}\defegal\Big\{ \util^{(\lambda)} :\RR^2 \to \RR
\,,\, \,\lambda\geq 0 \mid
\util^{(\lambda)}(x,\eta)=x+\lambda\big(-\rho(-x,\eta)+\gamma\big)
\Big\} \; .
\end{equation*}
\label{th:main}
\end{theoreme}

Table~\ref{tab:Risk-Utility} sums up parameterized  functions 
corresponding to usual risk measures.
 
\begin{table}[htbp]
\begin{center}
\begin{tabular}{c||c}
\hline
Risk measure $\Risk_{\rho}$ &
$\util^{(\lambda)}(x,\eta)$, $\lambda \geq 0$ \\[2mm]
\hline\hline
                  &                           \\
$\rho(-x,\eta) $ 
  &  $x-\lambda\rho(-x,\eta)+\lambda\gamma $  \\[2mm]
\emph{Variance} &  $x-\lambda(x+\eta)^2+\lambda\gamma $  \\[2mm]
\emph{Conditional Value-at-Risk}  &  $
x - \frac{\lambda}{1-p} (x+\eta)_- - \lambda
\eta +\lambda\gamma $ \\[3mm]
\emph{Weighted Mean Deviation} &
$x-\lambda\max\big\{-p(x+\eta),(1-p)(x+\eta)\big\}+\lambda\gamma$\\[3mm]
\emph{Optimized Certainty Equivalent} &
$x+\lambda {\util}(x+\eta) +\lambda \eta +\lambda\gamma $ \\
\hline
\end{tabular}
\end{center}
\caption{Usual risk measures and their corresponding family functions}
\label{tab:Risk-Utility}
\end{table}

Notice that the formulation~\eqref{eq:Eco} of Problem~\eqref{eq:Pb-ref} leads
us to introduce a new decision variable $\eta \in \RR$. This decision variable
comes from our choice of risk measure given by \eqref{eq:Risk-formulation}.

\bigskip

We shall now see that the formulation~(\ref{eq:Eco}) has connections
with the so called ``maxmin'' representation of \cite{Maccheroni:2002}, that we
briefly sketch.

Let us consider a continuous and convex weak order $\succcurlyeq$
over the set $\Delta Z$ of all lotteries
(simple probability distributions) defined on an outcome space $Z$.
The main result of \cite{Maccheroni:2002} is the following: if there exists a
best outcome\footnote{%
A best outcome for a preference relation $\succcurlyeq$ is an
element
$z^* \in Z$ such that $z^* \succcurlyeq r$ for all $r \in \Delta Z$.}
 and if decisions  are made independently of it, then there
exists a closed and convex set $\mathcal{\util}$ of utility functions
defined on $Z$ such that, for any lotteries $r$ and $q$, we have
\[
r \succcurlyeq q \Leftrightarrow
\min_{\util \in \mathcal{\util}} \int  \util(z) \mathrm{d} r(z)
\ge \min_{\util \in \mathcal{\util}} \int \util(z) \mathrm{d} q(z) \; .
\]
The interpretation
given by Maccheroni for this decision rule is the following one: a
conservative investor has an unclear evaluation  of the different
outcomes when facing lotteries. He then acts as if he were
considering many expected utility evaluations  and taking the worst
one\footnote{This reformulation is, in a sense, dual to the well known
form $\displaystyle\inf_{\PP\in\mathcal{P}}\int
\util(\wealth)\,\mathrm{d}\PP(\wealth)$, where $\mathcal{P}$ is a convex set of
probability measures (\cite{GilboaSchmeidler:1989}).}.

\subsection{Conditional Value-at-Risk and loss aversion}

Suppose that the risk constraint $\Risk_{\rho}$ is the Conditional
Value-at-Risk. The utility functions associated to this risk
measure are
\begin{equation*}
%\label{eq:Utility-CVaR}
\util^{(\lambda)}(x,\eta)=x - \frac{\lambda}{1-p} (-x-\eta)_+ - \lambda
\eta +\lambda\gamma\,.  
\end{equation*}
We consider only the $x$ argument (profit, benefit, etc.).
We interpret the $-\eta$ argument as an \emph{anchorage} parameter:
for $x\geq -\eta$, $x \mapsto \util^{(\lambda)}(x,\eta)=
x - \lambda \eta +\lambda\gamma$ has slope 1,
while it has slope 
\begin{equation*}
%\label{eq:loss-aversion-def}
\theta\defegal 1 + \frac{\lambda}{1-p}
\end{equation*}
for $x$ lower than $-\eta$, as in Figure~\ref{Utility-eEaR}. 
We interpret the parameter $\theta$ as a \emph{loss aversion} parameter
introduced by Kahneman and Tversky (see~\cite{KT92}).
Indeed, this utility function $x \mapsto \util^{(\lambda)}(x,\eta)$
expresses the property that one monetary unit than the anchorage $-\eta$
gives one unit of utility, while one unit less gives $-\theta$. 

\begin{figure}[htb]
\begin{center}
\input{UtilityCVaR.pstex_t} %\Figdir UtilityCVaR.pstex_t
\end{center}
\caption{Utility function attached to $\cvar$} 
\label{Utility-eEaR}
\end{figure}

\subsection{An illustration}
\label{sec:Portfolio_optimization_numerical_example}

Suppose that a DM splits its investment between 
a risk free asset $\xi^{0}$ (deterministic\footnote{%
Numerical value of 1~030\euro.}
and a risky asset $\xi^{1}$ 
(random following a Normal law $\mathcal{N}(M,\Sigma)$\footnote{%
Mean M=1~144~\euro\ and standard deviation $\Sigma$=249~\euro,
French stock market index.}, in
proportion $1-\acte \in [0,1]$), giving the value of the portfolio
\begin{equation*}
%\label{eq:richesse}
 \return\big(\acte,\xi\big)=\acte\xi^{0}+(1-\acte)\xi^1 =
\mu(\acte)+\sigma(\acte)N,\quad N\sim\mathcal{N}(0,1)  
\end{equation*}
with 
\begin{equation*}
 \mu(\acte)=\acte\xi^{0}+(1-\acte)M \text{and}
 \sigma(\acte)=(1-\acte)\Sigma\,.
\end{equation*}
The Conditional Value-at-Risk constraint on $-\return\big(\acte,\xi\big) $ is
\begin{equation*}
 \sigma(\acte)\cvar_{p}\big(-N\big)-\mu(\acte)\leq \gamma\,. 
%\label{eq:portfolio_Earning-at-Risk_constraint}
\end{equation*}
The portfolio maximization problem subject to
Conditional Value-at-Risk   risk constraint is:
\begin{subequations}
\label{eq:portfolio_maximization}
\begin{equation} 
\sup_{\acte\in[0,1]} \mu(\acte)
\end{equation}
\begin{equation}
 \sigma(\acte)\cvar_{p}\big(-N\big)-\mu(\acte)\leq \gamma\,. 
\end{equation}
\end{subequations}

By duality we find solutions of
problem~\eqref{eq:portfolio_maximization}:
\begin{equation*}
\acte\opt=\frac{M +\Sigma\cvar_{p}(-N)-\gamma}{M - \xi^0 +
\Sigma\cvar_{p}(-N)}\text{  and }
\lambda^{\opt}=\frac{1}{1 + \frac{\Sigma \cvar_{p}(-N)}{M - \xi^{0}}}\,.
\end{equation*}
Hence the optimal value of $\eta$ is $\eta\opt=\sigma(\acte\opt)
\VaR_{1-p}(N)-\mu(\acte\opt)$.

For two confidence levels $p=0.95$ and $p=0.99$, 
we exhibit the loss aversion parameter in Table~\ref{tab:theta_p}.
This latter takes high values, well above the empirical findings (median
value of 2.25 in~\cite{KT92}).

\begin{table}[t!]
\begin{center}
\begin{tabular}{||c|c|c|c||c|c|c|c||}
\hline
\hline
\multicolumn{4}{|l||}{ $p$=0.95 } & \multicolumn{4}{||l|}{ $p$=0.99 } \\[3mm]
\hline\hline
$\gamma$      & $\acte\opt$ & $\eta\opt$    & $\theta$  &
$\gamma$      & $\acte\opt$ & $\eta\opt$    & $\theta$  \\[3mm]
\hline\hline
-630~\euro   &   0          & -735   & 6.6 &
-496~\euro   &   0          & -565.1   & 22 \\[3mm]
\hline
-772.5~\euro  &   0.36      &  -839.5& 6.6 & 
-772.5~\euro  &   0.47      &  -785.9& 22 \\[3mm]
\hline
-978.5~\euro  &   0.87      & -991.9 & 6.6 & 
-978.5~\euro  &   0.84      & -955.6 & 22 \\[3mm]
\hline
-1030~\euro   &   1         & -1030  & 6.6 &
-1030~\euro   &   1         & -1030  & 22 \\[3mm]
\hline\hline
\end{tabular}
\end{center}
\caption{Loss aversion parameter with confidence levels
$p$=0.95 and $p$=0.99}
\label{tab:theta_p}
\end{table}

\section{Economic discussion}
\label{sec:Conclusion}

Our main result establishes a connection between some risk measures
and parameterized families of multi-attribute ``utility functions''.
We hope to be able to ``read'' some properties of the risk measure
from these latter.
For instance, focusing only on the $x$ argument in utility functions of
Table~\ref{tab:Risk-Utility},
we notice that the variance risk measure is associated to quadratic
utility functions. Now, the latter are well known for their poor
economic qualities (see \cite{Gollier:2001}), such as exhibiting
risk aversion increasing with wealth, for instance. 
In the $\cvar$ risk constraint case, 
the corresponding utility function is interpreted as loss aversion
utility function \emph{à la} Kahneman and Tversky (see~\cite{KT92}).

The role of the variable $\eta$ can also be discussed.
It is known that the optimal $\eta$ is the expectation of the optimal
profit in the variance case,  
and the Value-at-Risk in the Conditional Value-at-Risk.
Our formalism amounts to attributing a cost (disutility) to such a
variable $\eta$ when it is let loose, not necessarily fixed at its
optimal value. In the Optimized Certainty Equivalent case the optimal 
$\eta$ gives the optimal allocation between $\eta$ consumption and 
$(\return-\eta)$ investment.

\paragraph*{Acknowledgments.} We thank Alain Chateauneuf 
(Université Paris 1 Panthéon-Sorbonne) and 
Fabio Maccheroni (Università Bocconi, Milano) for fruitful discussions.  
We acknowledge Électricité de France for their funding.

\appendix
%\section{Appendix}

\section{Appendix: proof of the main result}
\label{sec:Appendix}

We shall show that~\eqref{eq:Pb-ref} is equivalent to
\begin{equation*}
%\label{eq:EquivEco}
\sup_{\acte\in\actes}\sup_{\eta\in\RR}\inf_{\lambda\in\RR_+}\EE\Big[
\return(\acte,\xi)-
\lambda\Big( \rho\big(-\return(\acte,\xi),\eta\big)-\gamma\Big) \Big]\, .
\end{equation*}
We suppose that all assumptions of Theorem~\ref{th:main}
are satisfied.

\subsection*{Equivalent Lagrangian formulation}

As is well known,
the Lagrangian associated to maximization problem~\eqref{eq:Pb-ref} is
\begin{equation*}
L(\acte,\lambda) \defegal \EE\big[\return(\acte,\xi)\big]-\lambda
\Big(\Risk_{\rho}\big(-\return(\acte,\xi)\big)-\gamma\Big) \, ,
\end{equation*}
where $\lambda\in\RR_{+}$ is a  Lagrange multiplier,
and we have~\eqref{eq:Pb-ref}
$\displaystyle \iff \sup_{\acte\in\actes}\inf_{\lambda \geq 0}L(\acte,\lambda)$.

We have:
\begin{eqnarray}
L(\acte,\lambda)&=&\EE\big[\return(\acte,\xi)\big]-\lambda \Big(
\inf_{\eta\in\RR}\EE\Big[\rho\big(-\return(\acte,\xi),\eta\big)\Big]
-\gamma\Big) \mtext{ by~\eqref{eq:Risk-formulation}} \nonumber\\
&=&\EE\big[\return(\acte,\xi)\big]-\lambda\Big
(-\sup_{\eta\in\RR}\EE\Big[-\rho\big(-\return(\acte,\xi),\eta\big)\Big]
-\gamma\Big)\nonumber\\
&=&\EE\big[\return(\acte,\xi)\big]+\lambda
\sup_{\eta\in\RR}\EE\Big[-\rho\big(-\return(\acte,\xi),\eta\big)\Big]
+\lambda\gamma\nonumber\\
&=&\sup_{\eta\in\RR}\Big(\EE\big[\return(\acte,\xi)\big]+\lambda
\EE\Big[-\rho\big(-\return(\acte,\xi),\eta\big)\Big]+\lambda \gamma\Big )\,,
\nonumber\\
&& \mtext{because $\lambda\geq 0$ and } \EE\big[\return(\acte,\xi)\big]
\mtext{ does not depend upon } \lambda
\nonumber\\
&=&\sup_{\eta\in\RR}\EE\Big[\return(\acte,\xi)-\lambda
\rho\big(-\return(\acte,\xi),\eta\big)+\lambda\gamma\Big]\,.
\nonumber
%\label{eq:Lbis}
\end{eqnarray}

Since~\eqref{eq:Pb-ref}
$\displaystyle \iff \sup_{\acte\in\actes}\inf_{\lambda \geq 0}L(\acte,\lambda)$,
it follows that
\begin{equation}
\label{eq:pbtrans} \eqref{eq:Pb-ref} \Leftrightarrow
\sup_{\acte\in\actes}\inf_{\lambda\in\RR_+}\sup_{\eta\in\RR}
\EE\Big[\return(\acte,\xi)-
\lambda\rho\big(-\return(\acte,\xi),\eta\big)+\lambda
\gamma\Big]\,.
\end{equation}
We now show that we can exchange
$\inf_{\lambda\in\RR_+}$ and $\sup_{\eta\in\RR}$\,.

\subsection*{Exchanging $\inf_{\lambda\in\RR_+}$ and $\sup_{\eta\in\RR}$}

Let $\acte\in\actes$ be fixed. Define $\Psi_{\acte} : \RR^2 \to \RR$ by
\begin{equation}
\label{eq:Psi_v}
\Psi_{\acte}(\lambda,\eta)\defegal
\EE\Big[\return(\acte,\xi)-\lambda\rho\big(-\return(\acte,\xi),\eta\big)
+\lambda\gamma\Big] \,.
\end{equation}
We can exchange $\inf_{\lambda\in\RR_+}$ and $\sup_{\eta\in\RR}$
in~(\ref{eq:pbtrans}) by the two following Lemmas.

\begin{lemma}
\label{lem:lemme1}   Let $\acte \in \actes$ be fixed.
If $\gamma < \Risk_{\rho}\big(-\return(\acte,\xi)\big) $, then
$\displaystyle\inf_{\lambda\in\RR_+}\,\sup_{\eta\in\RR}\,\Psi_{\acte}(\lambda,\eta)
=\displaystyle\sup_{\eta\in\RR}\,\inf_{\lambda\in\RR_+}\,\Psi_{\acte}(\lambda,\eta)
=-\infty$.
\end{lemma}

\begin{proof}

 By~\eqref{eq:Risk-formulation} and $\lambda \geq 0$, we have 
$\displaystyle\sup_{\eta\in\RR}\Psi_{\acte}(\lambda,\eta)=
\EE\big[\return(\acte,\xi)\big]-\lambda\big(\Risk_{\rho}\big(-\return(\acte,\xi)
\big)-\gamma\big)$.
Thus $\displaystyle\inf_{\lambda\in\RR_+}\sup_{\eta\in\RR}
\Psi_{\acte}(\lambda,\eta)=-\infty $, since
$ \gamma <\Risk_{\rho}\big(-\return(\acte,\xi)\big)$.

We always have $\displaystyle\sup_{\eta\in\RR}\inf_{\lambda\in\RR_+}
\Psi_{\acte}(\lambda,\eta)\leq\inf_{\lambda\in\RR_+}
\sup_{\eta\in\RR}\Psi_{\acte}(\lambda,\eta)$.

It holds that
$\displaystyle\sup_{\eta\in\RR}\inf_{\lambda\in\RR_+}
\Psi_{\acte}(\lambda,\eta)=-\infty $.

\end{proof}

The proof of the following Lemma is based on the Theorem 
hereafter (see~\cite[Chap. 2. Corollary 3.8]{Barbu:lagrange}).

\begin{theoreme}
\label{theo:psel}
Assume that $\Phi : \RR^p \times \RR^q \rightarrow
\RR\cup\{-\infty\}\cup\{+\infty\}$ is a convex-concave
\footnote{convex with respect to its first argument,
and concave with respect to its second argument.}
and l.s.c.-u.s.c.
\footnote{lower semicontinuous with respect to its first argument
 and upper semicontinuous with respect to its second argument.}
 mapping.
Assume that  $Y$ and $Z$ are two closed convex subsets
of \,$\RR^p$ and \,$\RR^q$ respectively, and that
there exists $(y^*,z^*) \in Y \times Z $ such that
\[
\left\{ \begin{array}{rcl}
\Phi(y^*,z) & \rightarrow -\infty, & \mtext{ when }
\|z\|\rightarrow +\infty \mtext{ and } z \in Z \\
\Phi(y,z^*) & \rightarrow +\infty, & \mtext{ when }
\|y\|\rightarrow +\infty \mtext{ and } y \in Y \, .
\end{array} \right.
\]
Then $\Phi$ admits a saddle point $(\bar{y},\bar{z}) 
\in Y \times Z$:
\[
\Phi(\bar{y},z)\leq
\Phi(\bar{y},\bar{z}) \leq \Phi(y,\bar{z}) \quad
\forall (y,z) \in Y \times Z \,.
\]
\end{theoreme}

\begin{lemma}
\label{lem:lemmme2} If $\gamma \geq \Risk_{\rho}\big(-\return(\acte,\xi)\big)$ then
$\Psi_{\acte}$ defined by~(\ref{eq:Psi_v}) admits a saddle point in
$\RR_+\times\RR$ and thus
$\displaystyle\sup_{\eta\in\RR}\inf_{\lambda\in\RR_+}\Psi_{\acte}(\lambda,\eta)=
\inf_{\lambda\in\RR_+}\sup_{\eta\in\RR}\Psi_{\acte}(\lambda,\eta) $.
\end{lemma}

\begin{proof}

Let $\eta^\star$ be such that
$\Risk_{\rho}\big(-\return(\acte,\xi)\big) =
\EE\big[\rho\big(-\return(\acte,\xi),\eta^\star \big)\big]$.
We have indeed supposed that
the infimum in~\eqref{eq:Risk-formulation} is achieved for any
$X=-\return(\acte,\xi)$ when $\acte$ varies in $\actes$.
We distinguish two cases.

\begin{description}
\item[a.] If $\gamma=\Risk_{\rho}\big(-\return(\acte,\xi)\big)$, any
  $(\lambda,\eta^\star)$ is a saddle point because \eqref{eq:Psi_v}
  gives
$\Psi_{\acte}(\lambda,\eta^\star)=\EE\big[\return(\acte,\xi)\big]$.

\item[b.] Assume now that $\gamma >
  \Risk_{\rho}\big(-\return(\acte,\xi)\big)$, and let us check the conditions
  of existence of a saddle point in Theorem~\ref{theo:psel}.

The function $\Psi_{\acte}(\lambda,\eta)=
\EE\Big[\return(\acte,\xi)\Big] - \lambda
\EE\Big[\rho\big(-\return(\acte,\xi),\eta\big) - \gamma\Big]$ is
\begin{itemize}
\item linear with respect to  $\lambda$ and thus convex in $\lambda$;
\item concave with respect to $\eta$ (the function : $\eta
\mapsto -\rho\big(-\return(\acte,\xi),\eta\big)$ is concave,
$\lambda\geq 0$ and the expectation operator preserves concavity).
\end{itemize}
Now, by assumption $\rho$ and $L_{\rho}(\Omega,\mathcal{F},\PP)$
satisfy assumption H2 with
$ \return(\acte,\xi) \in L_{\rho}(\Omega,\mathcal{F},\PP)$,
we have
\begin{itemize}
\item $\eta  \mapsto \EE\Big[\rho\big(-\return(\acte,\xi),\eta\big) - 
\gamma\Big]$
is continuous;
\item $\displaystyle\lim_{\eta \to +\infty} 
\EE\Big[\rho\big(-\return(\acte,\xi),\eta\big) - \gamma\Big]=+\infty$.
\end{itemize}
Thus, the function $\Psi_{\acte}$ is convex-concave, l.s.c.-u.s.c.
and satisfies
\[
\Psi_{\acte}(\lambda,\eta) \rightarrow -\infty,\, \mbox{when }
\eta \rightarrow +\infty \mtext{ for any } \lambda > 0 \, .
\]

Since $\gamma > \Risk_{\rho}\big(-\return(\acte,\xi)\big)=
\EE\big[\rho\big(-\return(\acte,\xi),\eta^\star \big)\big]$,
we have
\[
\displaystyle\Psi_{\acte}(\lambda,\eta^\star)=
 \EE\big[\return(\acte,\xi)\big]+\lambda\Big(\gamma -\Risk_{\rho} 
\big(-\return(\acte,\xi)\big)\Big)  \rightarrow +\infty,\, \mbox{when
$\lambda \rightarrow +\infty$ }\,.
\]
Hence, the function $\Psi_{\acte}$ admits a saddle point in 
$\RR_+\times\RR$.
\end{description}
\end{proof}

\end{document}